\documentclass[fleqn,usenatbib]{mnras}
\usepackage{newtxtext,newtxmath}
\usepackage[T1]{fontenc}

\DeclareRobustCommand{\VAN}[3]{#2}
\let\VANthebibliography\thebibliography
\def\thebibliography{\DeclareRobustCommand{\VAN}[3]{##3}\VANthebibliography}

\usepackage{orcidlink}
\usepackage{graphicx}
\usepackage{amsmath}
\usepackage{multirow}

\title[Mass-gap BH--NS Mergers]{Formation of Lower Mass-gap Black Hole--Neutron Star Binary Mergers through Super-Eddington Stable Mass Transfer}

\author[J.-P. Zhu et al.]{Jin-Ping Zhu\orcidlink{0000-0002-9195-4904},$^{1,2}$\thanks{E-mail: jin-ping.zhu@monash.edu}
Ying Qin\orcidlink{0000-0002-2956-8367},$^{3}$\thanks{E-mail:yingqin2013@hotmail.com}
Zhen-Han-Tao Wang\orcidlink{0000-0001-9592-6671},$^{4}$
Rui-Chong Hu\orcidlink{0000-0002-6442-7850},$^{5}$
Bing Zhang\orcidlink{0000-0002-9725-2524},$^{5,6}$\newauthor
Shichao Wu\orcidlink{0000-0002-9188-5435},$^{7,8}$
\\
$^{1}$School of Physics and Astronomy, Monash University, Clayton Victoria 3800, Australia\\
$^{2}$OzGrav: The ARC Centre of Excellence for Gravitational Wave Discovery, Australia\\
$^{3}$Department of Physics, Anhui Normal University, Wuhu, Anhui, 241002, China\\
$^{4}$Guangxi Key Laboratory for Relativistic Astrophysics, School of Physical Science and Technology, Guangxi University, Nanning 530004, China\\
$^{5}$Department of Physics and Astronomy, University of Nevada, Las Vegas, NV 89154, USA\\
$^{6}$Nevada Center for Astrophysics, University of Nevada, Las Vegas, NV 89154, USA\\
$^{7}$Max-Planck-Institut f{\"u}r Gravitationsphysik (Albert-Einstein-Institut), D-30167 Hannover, Germany\\
$^{8}$Leibniz Universit{\"a}t Hannover, D-30167 Hannover, Germany
}

\date{Accepted XXX. Received YYY; in original form ZZZ}
\pubyear{2023}

\begin{document}
\label{firstpage}
\pagerange{\pageref{firstpage}--\pageref{lastpage}}
\maketitle

\begin{abstract}

Super-Eddington accretion of neutron stars (NSs) has been suggested both observationally and theoretically. In this paper, we propose that NSs in close-orbit binary systems with companions of helium (He) stars, most of which systems form after the common-envelope phase, could experience super-Eddington stable Case BB/BC mass transfer (MT), and can sometimes occur {{accretion-induced collapse (AIC), resulting in the formation of}} lower mass-gap black holes (mgBHs). Our detailed binary evolution simulations reveal that AIC events tend to happen if the primaries NS have an initial mass $\gtrsim1.7\,M_\odot$ with a critical accretion rate of $\gtrsim300$ times the Eddington limit. These mgBHs would have a mass nearly equal to or slightly higher than the NS maximum mass. The remnant mgBH--NS binaries after the core collapses of He stars are potential progenitors of gravitational-wave (GW) source. Multimessenger observation between GW and kilonova signals from a population of high-mass binary NS and mgBH--NS mergers formed through super-Eddington stable MT are helpful in constraining the maximum mass and equation of state of NSs. S230529ay, a mgBH--NS merger candidate recently detected in the fourth observing run of the LIGO-Virgo-KAGRA Collaboration, could possibly originate from this formation scenario.

\end{abstract}


\section{Introduction}
On the one hand, the X-ray and radio observations of Galactic pulsars revealed a likely NS maximum mass of $\sim2-2.3\,M_\odot$ \citep[e.g.,][]{Antoniadis2013,Alsing2018,Romani2022}, consistent with the maximum mass inferred by the observations of NSs in GW binaries \citep[e.g.,][]{Margalit2017,Landry2021,Zhu2022Population,Abbott2023Population}. On the other hand, the measurements of the mass distribution of BHs in Galactic X-ray binaries suggested a lower boundary close to $\sim5\,M_\odot$ \citep{Bailyn1998,Ozel2010,Farr2011}. This led to the conjecture of the presence of the mass gap between the heaviest NSs and lightest BHs, which was thought to be potentially caused by the observational bias \citep{Kreidberg2012}, natal kick \citep{Mandel2021Binary}, or the intrinsic physics mechanism of core-collapse supernova (SN) explosions, e.g., the rapid model \citep{Fryer2012}. However, recent electromagnetic (EM) observations discovered a few compact objects with mass plausibly lying in the range of $\sim2.5-5\,M_\odot$, indicating that the putative mass gap could be at least partly populated. For example, gravitational microlensing identified eight compact-object candidates with masses within the mass gap \citep{Wyrzykowski2020}. \cite{Thompson2019} reported that the unseen companion of giant star 2MASS J05215658+4359220 could have a mass of $3.3_{-0.7}^{+2.8}\,M_\odot$, which could be a noninteracting mgBH. {{Using radial velocity measurements, \cite{Rivinius2020} suggested HR\,6819, a star in a hierarchical triple system, might be accompanied by an unseen companion, which could be a nonaccreting BH with a mass of $\geq4.2\,M_\odot$. \cite{vanderMeij2021} calculated a mass of 4U\,1700-37 of $2.54\,M_\odot$, which could either be a NS or a BH in the mass gap. \cite{Andrews2022} used Gaia data release to estimate the masses of the dark companions in wide-orbit binaries, which span a range of $1.35-2.7\,M_\odot$, partially intersecting with the mass gap. }}

{{Currently, it remains uncertain that whether X-ray binaries and GW sources follow a similar evolutionary pathway \citep[e.g.,][]{Belczynski2021,Fishbach2022}.}} It is also expected that GW searches could discover a number of mgBHs in the merging systems of binary BH--NS and binary BH (BBH). During the third observing run (O3) of the LIGO-Virgo-KAGRA (LVK) Collaboration, a few GW candidates that potentially contain a mass-gap compact object were indeed detected. GW190814 was reported to be a merger between a $23.2^{+1.1}_{-1.0}\,M_\odot$ BH and a mass-gap compact object with a mass of $2.59^{+0.08}_{-0.09}\,M_\odot$ \citep{Abbott2020GW190814}. The LVK Collaboration also detected a similar, but marginal event namely GW200210\_092254 in O3 \citep{Abbott2021GWTC3}, where the component masses were inferred to be $24.1^{+7.5}_{-4.6}\,M_\odot$ and $2.83^{+0.47}_{-0.42}\,M_\odot$, respectively. The posterior distributions for the primary source mass of the low-significance BH--NS candidate GW190426\_152155 and the BH--NS merger GW200115, i.e., $5.7^{+3.9}_{-2.3}\,M_\odot$ and $5.9^{+1.4}_{-2.1}\,M_\odot$, partly lie in the mass gap \citep{Abbott2021GWTC2,Abbott2021BHNS}. However, by applying alternative astrophysically motivated priors, the primary mass of GW200115 would be more tightly constrained to be $7.0^{+0.4}_{-0.4}\,M_\odot$ \citep{Mandel2021}, which could be completely higher than the mass gap. Furthermore, the secondary component of GW191113\_071753 has a mass of $5.9^{+4.4}_{-1.3}\,M_\odot$ with $13\%$ probability of being a mgBH \citep{Abbott2021GWTC3}. Despite the detection of a few GW candidates containing mass-gap compact objects in O3, the population properties of merging compact binaries using GWs in GWTC-2 and GWTC-3 still indicated a relative dearth of events with masses in the mass gap \citep{Farah2022,Zhu2022Population,vanSon2022,Olejak2022,Ye2022,Biscoveanu2023,Abbott2023Population}. 

Systematic investigations of BH--NS systems formed through isolated binary evolution have been recently explored \cite[e.g.,][]{Jaime2021,Hu2022BHNS,Xing2023,Wang2024}. Binary population synthesis showed that {{the rapid SN model does not form any mgBHs}} \citep[e.g.,][]{Belczynski2012,Giacobbo2018}, while the delayed and stochastic models \citep{Mandel2020} suggested that $\sim30-80\%$ of BH--NS mergers and $\sim20-40\%$ of BBH mergers are expected to host at least one mgBH \citep[e.g.,][]{Shao2021,Drozda2022}. Both scenarios appear to be inconsistent with the observational results from GWTC-3. On the other hand, NSs can occur AIC to mgBHs via accretion when NSs grow to the point of exceeding the maximum mass allowed by their equation of state (EoS). AICs of accreting NSs are mostly predicted to take place in intermediate/low-mass X-ray binaries with companions of degenerated hydrogen/He dwarf stars \citep{MacFadyen2005,Dermer2006,Giacomazzo2012,Gao2022,Chen2023}. Furthermore, based on the standard scenario of the formation of a binary NS (BNS) system \citep[e.g.,][]{Bhattacharya1991,Tauris2017,VignaGomez2018}, an NS orbiting a main-sequence star needs to undergo common-envelope phase to become a close-orbit NS--He star system when the secondary star expands to a giant star. The NS could accrete materials from the hydrogen envelope {{during the common-envelope phase}} and from the He star during the subsequent Case BB/BC MT phase to increase its mass. {{As predicted by several simulations \citep[e.g.,][]{MacLeod2015,Esteban2023}, since the common-envelope stages are months-long with accretion rates up to $\dot{M}\lesssim0.1\,M_\odot$ if the super-Eddington accretion for the NS is permissible, the mass transfer (MT) during the common-envelope phase is usually limited during this phase.}} The discovery that a number of ultra-luminous X-ray sources (ULXs) are pulsating NSs unambiguously suggested that {{a fraction of}} NSs {{in binary systems}} can accrete at a super-Eddington accretion rate \citep[see][for a review]{Kaaret2017}. Population synthesis simulations by \cite{Shao2019} suggested that a significant fraction of NS ULXs in a Milky Way–like galaxy could contain a He star companion. Most recently, \cite{Zhou2023} for the first time identified a He donor star in NGC 247 ULX-1. Thus, some NSs could experience super-Eddington accretion in an NS--He star binary during the stable MT stage. {{The MT rates could range from a few $10^{-5}\,M_\odot\,{\rm yr}^{-1}$ to a few $10^{-4}\,M_\odot\,{\rm yr}^{-1}$ with durations of a few $10^4\,{\rm yr}$. Thus, AIC of an NS to a mgBH could happen if the NS reaches its maximum mass after accreting sufficient materials.}} {{In this paper, we present a new formation scenario in which an accreting NS during the stable Case BB/BC MT with super-Eddington accretion could undergo AIC leading to the formation of an mgBH.}} The final remnant system could be an ideal GW source of an mgBH--NS binary merger if the binary system survives the second SN.

\section{Modelling} \label{sec:model}

\subsection{Super-Eddington Accretion of NSs}

Traditionally, the accretion rate of compact objects is thought to be limited by the Eddington limit, i.e.,
\begin{equation}
\begin{split}
    &\dot{M}_{\rm Edd} = \frac{4\pi GM_{\rm acc}}{\kappa c\eta}\\
    &= 3.6\times10^{-8}\left(\frac{M_{\rm acc}}{1.4\,M_\odot}\right)\kappa_{-0.47}^{-1}\eta_{-1}^{-1}\,M_\odot\,{\rm yr}^{-1},
\end{split}
\end{equation}
with the Eddington luminosity of
\begin{equation}
\begin{split}
    &L_{\rm Edd} = \eta\dot{M}_{\rm Edd}c^2 = \frac{4\pi GM_{\rm acc}c}{\kappa} \\
    &=2.1\times10^{38}\left(\frac{M_{\rm acc}}{1.4\,M_\odot}\right)\kappa_{-0.47}^{-1}\,{\rm erg}\,{\rm s}^{-1},
\end{split}
\end{equation}
where $M_{\rm acc}$ is the accretor's mass, $\kappa$ is the opacity of the accreting material, $G$ is the gravitational constant, and $c$ is the speed of light. {{Hereafter, the conventional notation $Q_x = Q/10^x$ is adopted in cgs units.}} The radiation efficiency of accretion $\eta$ is set to be $\eta = 0.1$ for an NS accretor and $\eta = 1-\sqrt{1-({M_{\rm BH}/M_{\rm BH,i}})^2/9}$ if $M_{\rm BH}<\sqrt{6}M_{\rm BH,i}$ for a BH accretor \citep{Podsiadlowski2003}, where $M_{\rm BH}$ is the BH mass, and {{$M_{\rm BH,i}$ is the initial BH mass which is equal to the NS maximum mass $M_{\rm NS,max}$}}. In our work, $M_{\rm NS,max}\sim2.2\,M_\odot$ is defined following the constraints by the observations of Galactic pulsars and GW binaries \citep[e.g.,][]{Margalit2017,Romani2022,Zhu2022Population,Abbott2023Population}.

The discoveries of NS ULXs demonstrated that NSs can accrete at super-Eddington rate. Some of ULXs, e.g., M82 X-2 \citep{Bachetti2014}, NGC 7793 P13 \citep{Furst2016}, and NGC 1313 X-2 \citep{Sathyaprakash2019} have an apparent luminosity of a few $10^{40}\,{\rm erg}\,{\rm s}^{-1}$ implying that the accretion rate could be $\gtrsim100\,\dot{M}_{\rm Edd}$. \cite{Israel2017} reported that NGC 5907 ULX-1 and some other extreme ULXs that might harbor NSs can even have a maximum X-ray luminosity of a few $10^{41}\,{\rm erg}\,{\rm s}^{-1}$ corresponding to $\sim1000\,\dot{M}_{\rm Edd}$ {{if the emission is isotropic}}. 

For {{a standard gas-pressure-dominated thin disk with super-Eddington accretion}}, the location at which {{the accretion disk just possesses a local luminosity reaching the Eddington limit and has mass loss}} is defined as the spherization radius \citep{Shakura1973}, i.e., 
\begin{equation}
R_{\rm sph} \approx \frac{3\kappa\dot{M}}{8\pi c},
\end{equation}
where $\dot{M}$ is the accretion rate of the disk. Furthermore, {{the NS magnetic field can interact with the accretion disk.}} The disk around a magnetized NS is disrupted at the magnetospheric radius {{\citep{Frank2002}}}, which is given by 
\begin{equation}
    R_{\rm mag}=\xi R_{\rm A} = \xi\left(\frac{\mu^2}{\dot{M}_{\rm in}\sqrt{2GM_{\rm NS}}}\right)^{2/7},
\end{equation}
where $R_{\rm A}$ is the Alfv$\acute{\rm e}$n radius, $\xi$ is the dimensionless coefficient \citep{Ghosh1979,Wang1996,Long2005,Kulkarni2013}, $\dot{M}_{\rm in}$ is the accretion rate at $R_{\rm mag}$, $M_{\rm NS}$ is the NS mass, $\mu=B_{\rm NS}R_{\rm NS}^3$ is the magnetic moment of the NS with $B_{\rm NS}$ and $R_{\rm NS}$ being the magnetic field strength and radius of the NS, respectively. One can determine a critical accretion rate of $\dot{M}_{\rm cr}\propto\mu^{4/9}$ {{which provides a limit on $\dot{M}_{\rm in}$}} by equating $R_{\rm mag}$ with $R_{\rm sph}$. Thus, the NS accretion with a higher super-Eddington critical rate requires a larger NS magnetic field. If $R_{\rm mag} \gtrsim R_{\rm sph}$, {{for a subcritical disk}}, we have $\dot{M}_{\rm in} = \dot{M}$; otherwise, {{for a supercritical disk,}} then $\dot{M}_{\rm in} = \dot{M}(R_{\rm mag}/R_{\rm sph})=\dot{M}_{\rm cr}$.

The model of \cite{Shakura1973} is based on the assumption of a standard geometrically thin disk. Most recently, \cite{Chashkina2017} proposed that the NS accretion disk could be {{geometrically thick and radiation-pressure-dominated in its inner part}}, {{leading to a larger magnetosphere size and, hence, a larger}} critical accretion rate of 
\begin{equation}
\label{equ:SuperEddRate1}
    \dot{M}_{\rm cr,1} \simeq 35\,\alpha_{-1}^{2/9}\mu_{30}^{4/9}\dot{M}_{\rm Edd}.
\end{equation}
For a higher accretion rate, {{the inner disk would then become advection-dominated}} and the critical rate can be further enhanced to \citep{Chashkina2019}
\begin{equation}
\label{equ:SuperEddRate2}
    \dot{M}_{\rm cr,2} \simeq 200\,\alpha_{-1}^{2/9}\mu_{30}^{4/9}\dot{M}_{\rm Edd},
\end{equation}
{{Therefore, one value of $\dot{M}_{\rm in}$ can be determined by two different magnetic field strengths. For example, for a $1.4\,M_\odot$ NS with an accretion rate of $100\,\dot{M}_{\rm Edd}$, the magnetic field strength can be $\sim5.6\times10^{12}\,{\rm G}$ if the disk is radiation-dominated by Equation (\ref{equ:SuperEddRate1}) or $\sim1.1\times10^{11}\,{\rm G}$ if the disk is advection-dominated by Equation (\ref{equ:SuperEddRate2}), respectively. Here, $\alpha=0.1$ and an NS radius of $R_{\rm NS}=12.5\,{\rm km}$ are adopted \citep[e.g.,][]{Miller2019,Landry2021}. Since we will discuss the influence of magnetic field decay on accretion rates in Section \ref{sec:properties} and two different magnetic field strengths will give consistent results, for simplicity, }} we assume that the inner disk of the super-Eddington accretion disk during the Case BB/BC MT stage could always be advection-dominated and, hence, we can use Equation (\ref{equ:SuperEddRate2}) to estimate critical accretion rate for specific NS magnetic field. {{The accretion rate of mgBHs formed after AICs of NSs is assumed to be limited by the Eddington limit, i.e., $\dot{M}_{\rm cr} = \dot{M}_{\rm Edd}$, because the magnetic field effect stops operating. The mass accretion rate of the compact object, including NS and BH, can be expressed as}}  
\begin{equation}
\dot{M}_{\rm CO}=\begin{cases}(1-\eta)\dot{M} & {\rm if}~\dot{M}<\dot{M}_{\rm cr},\\(1-\eta)\dot{M}_{\rm cr}, & {\rm if}~\dot{M}>\dot{M}_{\rm cr}.\end{cases}
\end{equation}

\subsection{Physics Implemented in MESA} \label{sec:MESA}

We adopt the release version \texttt{mesa-r15140} of the Modules for Experiments in Stellar Astrophysics (\texttt{MESA}) stellar evolution code \citep{Paxton2011,Paxton2013,Paxton2015,Paxton2018,Paxton2019,Jermyn2023} to evolve secondary He stars, which are assumed to be tidally locked by primary NS companions {{(treated as point masses)}}. The {{zero-age He main-sequence}} stars are created following the same method with \cite{Qin2018,Qin2023,Bavera2020,Bavera2021,Hu2022BHNS,Hu2023,Fragos2023,lv2023,Zhang2023,Wang2024}, and then are relaxed to reach the thermal equilibrium. Convection is implemented based on the Ledoux criterion, and mixing-length theory \citep{bohmvitense1958} with a mixing length of $\alpha_{\rm{MLT}}=1.93$ is adopted. Convective mixing is treated as a top step decay process with an overshoot parameter $l_{\rm{ov}}=0.1H_{\rm{p}}$, where $H_{\rm{p}}$ is the pressure scale height at the Ledoux boundary limit. We also consider semiconvection \citep{langer1983} with an efficiency parameter $\alpha_{\rm{SC}}=1.0$ in our model. The network of \texttt{approx12.net} is chosen for nucleosynthesis.

For stellar winds, we use the ``\texttt{Dutch}" scheme for both \texttt{RGB} and \texttt{AGB} phase, as well as the cool and hot wind. We adopt the default \texttt{RGB\_to\_AGB\_to\_wind\_switch = 1d-4}, as well as \texttt{cool\_wind\_full\_on\_T = 0.8d4} and \texttt{hot\_wind\_full\_on\_T = 1.2d4}. The standard ``\texttt{Dutch}" scheme was calibrated by multiplying with a Dutch scaling factor of 0.667 to match the recently updated modeling of He star wind mass loss \citep{Higgins2021} is adopted. Angular momentum transport and rotational mixing diffusive processes \citep{heger2000i,Heger2000ii}, including the effects of Eddington-Sweet circulations, and the Goldreich-Schubert-Fricke instability, as well as secular and dynamical shear mixing, are incorporated. We adopt diffusive element mixing from these processes with an efficiency parameter of $f_c = 1/30$ \citep{Chaboyer1992,Heger2000ii}. Additionally, the traditional Spruit-Tayler \citep{spruit2002} dynamo-induced angular momentum transport is implemented in our models. 

We model MT following the Kolb scheme \citep{Kolb1990}, and the implicit MT method \citep{Paxton2015} is adopted. The dynamical tides are applied to massive stars with radiative envelopes, and the corresponding timescale for orbital synchronization is calculated following the prescription in \cite{Hurley2002}. We adopt the updated tidal torque coeﬃcient $E_2$ provided in \cite{Qin2018}.

\section{Properties of mgBH--NS Formed through Super-Eddington Accretion}
\label{sec:properties}

\begin{figure}
    \centering
    \includegraphics[width = 1\linewidth , trim = 66 34 93 56, clip]{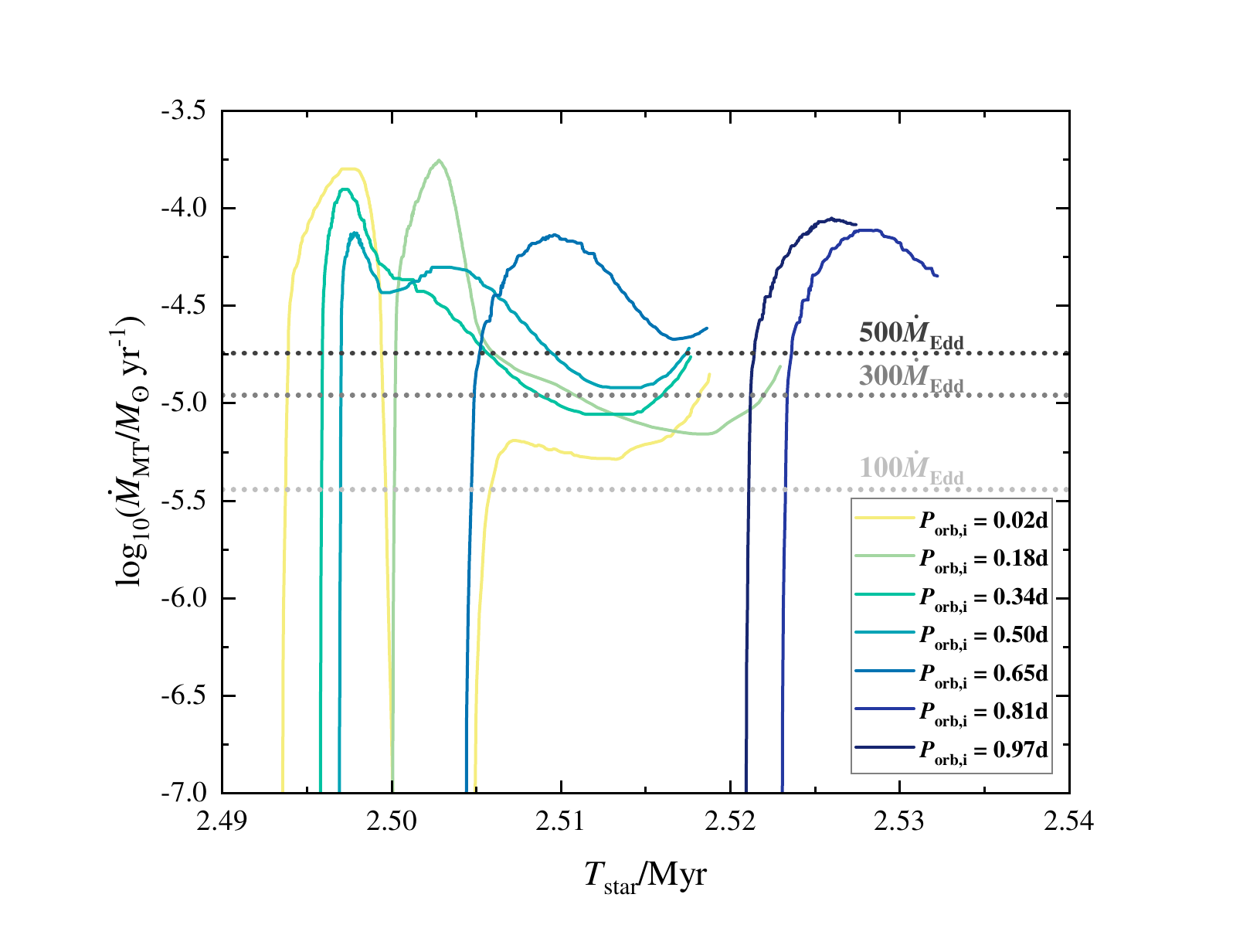}
    \caption{{{MT rate of binary systems as a function of the star age. We set an initial NS mass of $M^{\rm NS}_{\rm 1,i}=1.4\,M_\odot$, an initial He star mass of $M^{\rm He}_{\rm 2,i}=2.5\,M_\odot$ and $Z=Z_\odot$. Seven different initial orbital periods of $P_{\rm orb,i}=0.02,\,0.18,\,0.34,\,0.50,\,0.65,\,0.81,\,{\rm and} 0.97\,{\rm d}$ are considered. Three different super-Eddington accretion rates for a $1.4\,M_\odot$ NS, including $100\,\dot{M}_{\rm Edd}$, $300\,\dot{M}_{\rm Edd}$, and $500\,\dot{M}_{\rm Edd}$, are marked with dashed lines. }}}
    \label{fig:MT}
\end{figure}

We choose three different initial masses for the primary NS, including $M^{\rm NS}_{\rm 1,i} = 1.4,$ $1.7,$ and $2.0\,M_\odot$. If the primary NS mass exceeds $M_{\rm NS,max}$ during the process of super-Eddington accretion, the AIC to mgBH can occur. The initial He star mass $M^{\rm He}_{\rm 2,i}$ is in a range of $2.5-8\,M_{\odot}$. We follow the evolution of these He stars from the {{zero-age He main-sequence}} until carbon depletion occurred in their cores. {{Then, we estimate the baryonic mass of the remnant NS and BH formed through core collapse based on the delayed SN mechanism rather than the rapid SN mechanism \citep{Fryer2012}, prompted by recent plausible discoveries on compact objects in the lower mass gap \citep[e.g.,][]{Wyrzykowski2020,Thompson2019,Abbott2020GW190814}. The baryonic mass of the remnant NS formed through electron-capture SN is set to $1.38\,M_\odot$ following \cite{Fryer2012} if the pre-SN carbon-oxygen core mass is within the range between $1.37-1.43\,M_\odot$ \citep{Tauris2015}.}}  We also take into account neutrino loss as in \cite{Zevin2020}. For the binary systems which could have MT, we cover the initial orbital periods $P_{\rm orb,i}$ of $0.04 - 40\,{\rm d}$. Three critical super-Eddington accretion rates of $\dot{M}_{\rm cr}=100\,\dot{M}_{\rm Edd}$, $300\,\dot{M}_{\rm Edd}$ and $500\,\dot{M}_{\rm Edd}$ are considered. For a $1.4\,M_\odot$ NS, these three rates correspond to $\sim3.6\times10^{-6}\,\kappa_{-0.47}^{-1}\,M_\odot\,{\rm yr}^{-1}$, $\sim1.1\times10^{-5}\,\kappa_{-0.47}^{-1}\,M_\odot\,{\rm yr}^{-1}$ and $\sim1.8\times10^{-5}\,\kappa_{-0.47}^{-1}\,M_\odot\,{\rm yr}^{-1}$, which are always much lower than the MT rate of Case BB/BC (i.e., from a few $10^{-5}\,M_\odot\,{\rm yr}^{-1}$ to a few $10^{-4}\,M_\odot\,{\rm yr}^{-1}$; {{see Figure \ref{fig:MT}}}). By setting $\alpha=0.1$ and $R_{\rm NS} = 12.5\,{\rm km}$ \citep[e.g.,][]{Miller2019,Landry2021} in Equation (\ref{equ:SuperEddRate2}), NSs are required to have magnetic field of $\sim1.1\times10^{11}\,{\rm G}$, $\sim1.3\times10^{12}\,{\rm G}$, and $\sim4.0\times10^{12}\,{\rm G}$, respectively. {{Besides defined parameters for evolving binary systems introduced in Section \ref{sec:MESA}, we set the grids of $M^{\rm NS}_{1,i}$, $M^{\rm He}_{\rm 2,i}$, $P_{\rm orb,i}$, metallicity $Z$, and $\dot{M}_{\rm cr}$ as initial input parameters to simulate the final primary mass $M_{\rm 1,f} $and second-born NS/BH mass $M_{\rm 2,f}$ through \texttt{MESA}. When we consider the influence of accretion-induced magnetic field on the accreted mass and, hence, the final primary mass, we transfer the critical accretion rates to magnetic field strengths by Equation (\ref{equ:SuperEddRate2}) and evolve them with accreted masses. }}

\begin{figure*}
     \centering
     \includegraphics[width=1.0\textwidth]{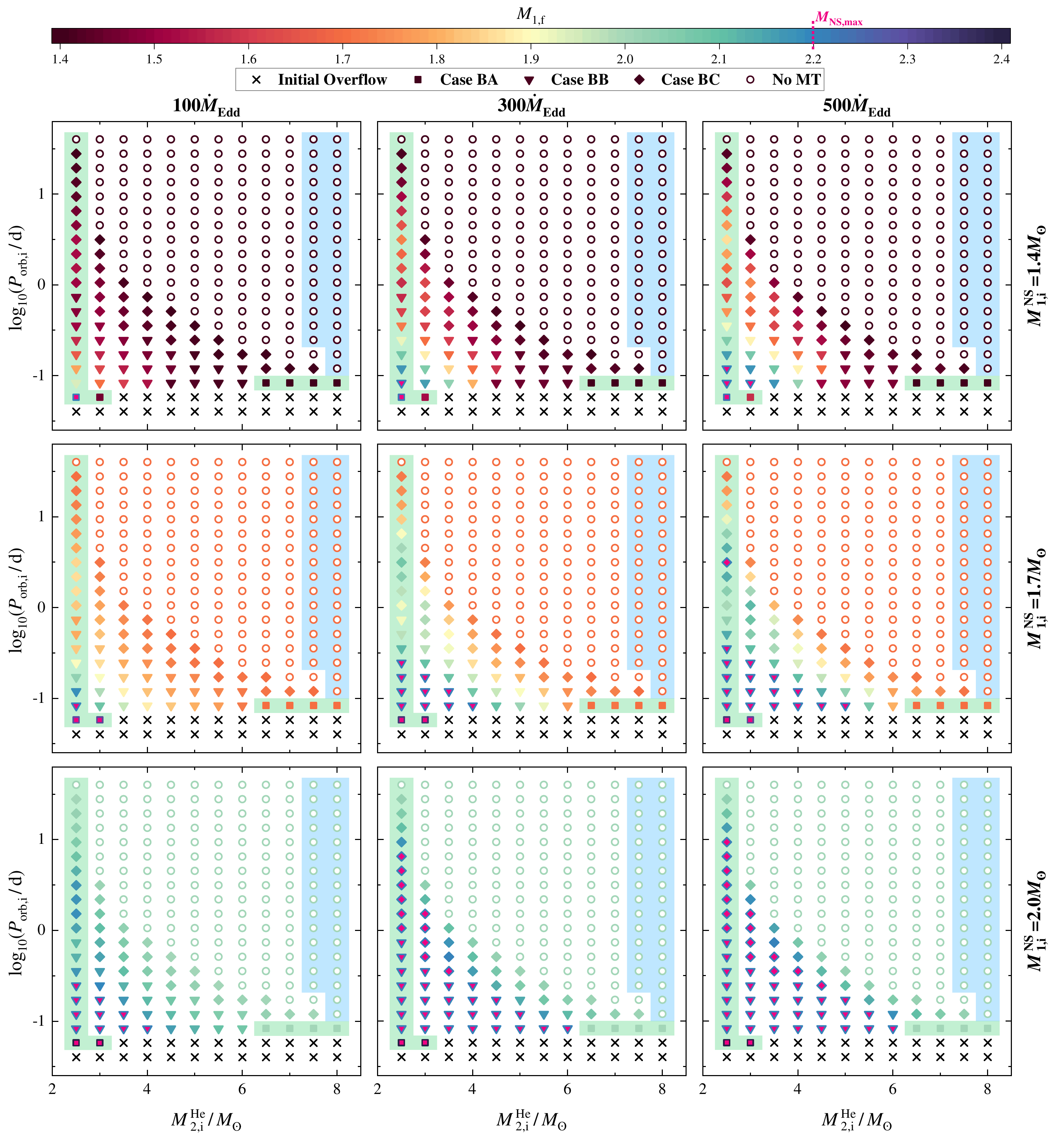}
     \caption{{{Final remnant mass}} (the color bars) as a function of the initial mass of secondary He star $M_{\rm 2,\rm i}^{\rm He}$, and orbital period $P_{\rm orb,i}$. Three different $M_{1,\rm i}^{\rm NS}$, including $1.4\,M_\odot$ (top panels), $1.7\,M_\odot$ (middle panels), and $2.0\,M_\odot$ (right panels) are considered. For each $M_{1,\rm i}^{\rm NS}$, we set three different critical accretion rates, i.e., $100\,\dot{M}_{\rm Edd}$, $300\,\dot{M}_{\rm Edd}$, and $500\,\dot{M}_{\rm Edd}$ from left to right panels. Black crosses, squares, triangles, diamonds, and circles represent the conditions of initial overflow, Case BA, Case BB, Case BC, and no MT, respectively. {{The points with pink interiors}} mark systems that can occur AIC of NS. {{Blue and green shading indicates where the He stars end up as WDs and BHs, respectively.}} We note that the color bars are different for each $M_{1,\rm i}^{\rm NS}$.}
     \label{fig:AccretedMass_Zsun}
\end{figure*} 

\begin{figure*}
     \centering
     \includegraphics[width=0.667\textwidth]{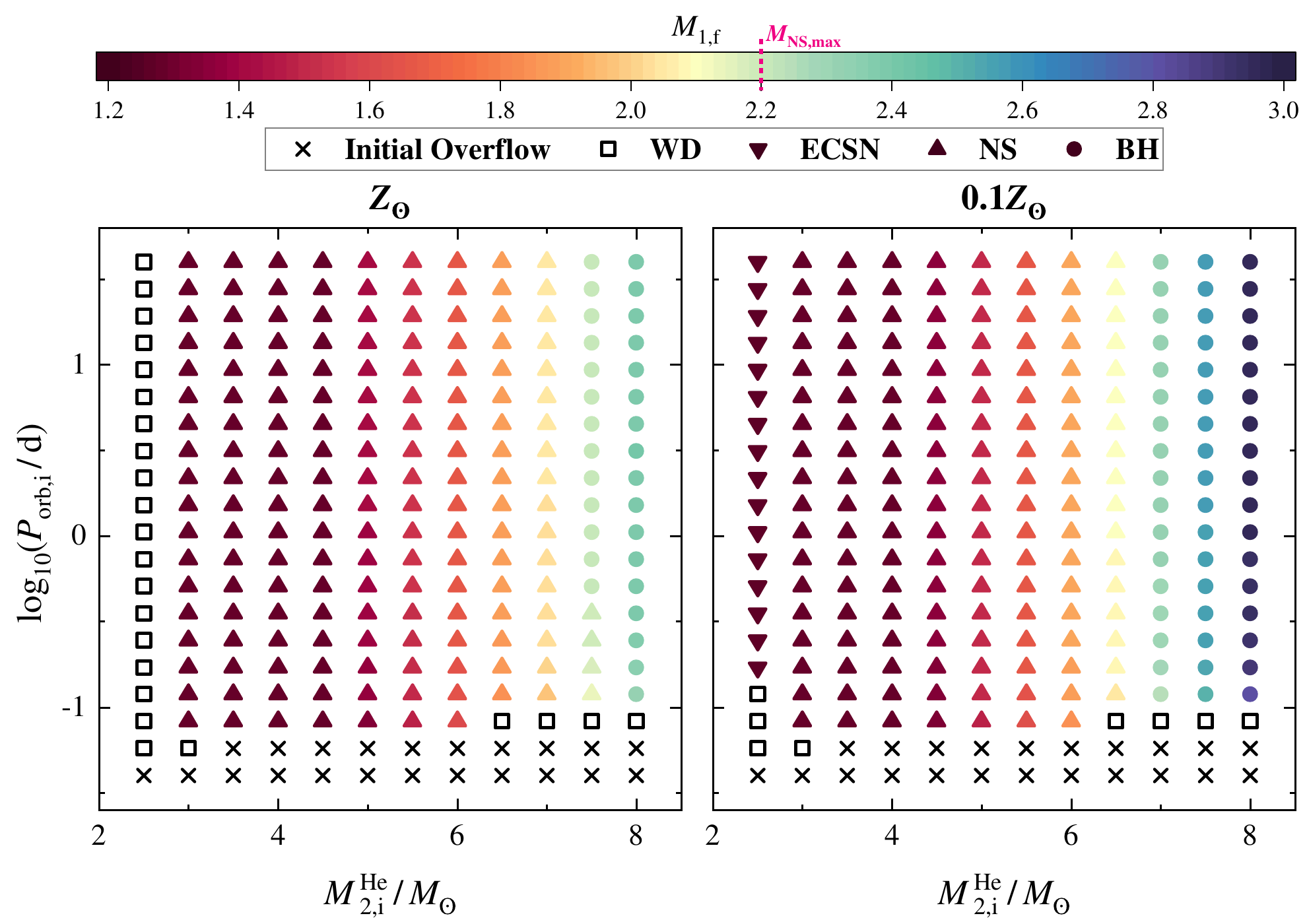}
     \caption{Mass of second-born NS and BH (the color bars) as a function of the initial mass of secondary He star $M_{2,\rm i}^{\rm He}$ and initial orbital period $P_{\rm orb,i}$. Left and right panels represent two different metallicities, i.e., $Z_\odot$ and $0.1\,Z_\odot$, respectively. Squares, lower triangles, upper triangles, and circles are the outcomes of white dwarf, NS formed through electron-capture SN, NS formed through core-collapse SN, and BH. Since different initial primary NS masses and Eddington accretion limits have little impact on the mass of resulting compact objects, $M^{\rm NS}_{1,i}=1.4\,M_\odot$ and an accretion limit of $100\dot{M}_{\rm Edd}$ are set here, while results by considering other different parameters limits are not shown. {{The maximum NS mass is set to $M_{\rm NS,max}=2.2\,M_\odot$.}} }
     \label{fig:NSMass}
\end{figure*} 

{{We firstly ignore the possible decay of NS magnetic field and, hence, the accreting NSs can always have a constant critical super-Eddington accretion rate, since it is currently still unclear which physical effect, such as spindown-induced flux expulsion, Ohmic evolution of the crustal field, and diamagnetic screening of the field by accreted plasma, dominates the decay of magnetic field \citep[e.g.,][]{Bhattacharya2002}.}} It is worth noting that the observation of a large bubble nebula surrounding NGC 1313 X-2 suggested that this ULX pulsar could have maintained a super-Eddington phase for more than $1\,{\rm Myr}$, indicating that the magnetic field of this source has not been suppressed during the accreting phase \citep{Sathyaprakash2019}. Furthermore, despite the experience of extensive MT, some long-history accreting-powered pulsars could still have a strong magnetic field. For instance, the NS in low-mass X-ray binary 4U 1626-67 is almost older than $100\,{\rm Myr}$, but its magnetic field is still as strong as a few $10^{12}\,{\rm G}$ \citep{Verbunt1990}. Therefore, some NSs could accrete a large abundance of materials without losing their strong magnetic field.

Figure \ref{fig:AccretedMass_Zsun} displays the parameter space that allows the formation of mgBH--NS binaries in a solar-metallicity environment, where $Z_\odot = 0.0142$ \citep{Asplund2009} is employed in this work. We find that mgBH--NS binaries can hardly be formed via super-Eddington accretion from binary systems composed of $\lesssim1.4\,M_\odot$ NSs and He stars. When the primary NSs with $M_{1,\rm i}^{\rm NS}\sim1.7\,M_\odot$ have an accretion rate of $\gtrsim300\,\dot{M}_{\rm Edd}$ or the NSs with $M_{1,\rm i}^{\rm NS}\sim2.0\,M_\odot$ have an accretion rate of $\sim100\,\dot{M}_{\rm Edd}$, AICs can happen only if the He star companions have an initial mass of $M_{\rm 2,i}^{\rm He}\sim3-5\,M_\odot$ and the binary systems have extremely close orbits with $P_{\rm orb,i}\lesssim0.2\,{\rm d}$. {{These AICs majorly occur duration the Case BB MT, and the second-born NSs could have a mass of $M_{2}^{\rm NS}\sim1.1-1.5\,M_\odot$ (see the left panel of Figure \ref{fig:NSMass}). Furthermore, all NSs with a mass range of $\gtrsim2.0\,M_\odot$ and an accretion rate of $\gtrsim300\,\dot{M}_{\rm Edd}$ that undergo the Case BB MT, as well as some undergoing the Case BC MT, can easily experience AIC to BHs. If the secondary is a $\sim3\,M_\odot$ He star, the formation of mgBHs can be allowed for binary systems with an initial orbital period reaching $P_{\rm orb,i}\sim2\,{\rm d}$. In binary systems with $P_{\rm orb,i}\lesssim0.1\,{\rm d}$, the primary NSs can finally collapse, even if the secondary He stars are as massive as $M_{\rm 2,i}^{\rm He}\sim6\,M_\odot$. We note that the NS could undergo AIC if the initial binary system is composed of a $1.7\,M_\odot$ NS and a $2.5\,M_\odot$ He star that end up as a white dwarf (WD) with a critical accretion rate of $300\,\dot{M}_{\rm Edd}$ (see the right-middle panel in Figure \ref{fig:AccretedMass_Zsun}) while AICs in the nearby \texttt{MESA} grid systems are prohibited, because the duration of the MT rate of this system above the rate is much longer compared to those of nearby \texttt{MESA} grid systems, as shown in Figure \ref{fig:MT}. }}

\begin{figure*}
     \centering
     \includegraphics[width=1.0\textwidth]{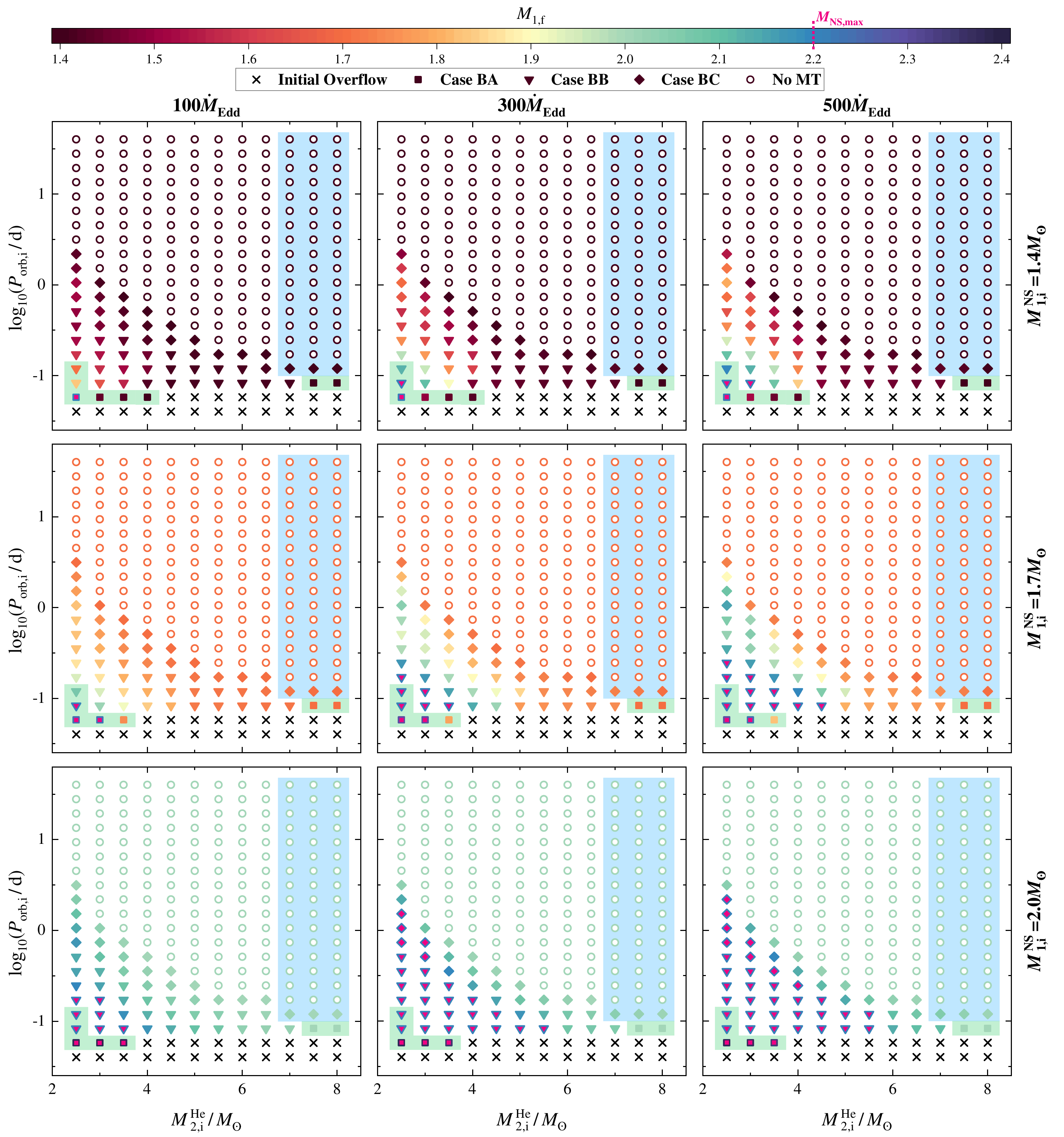}
     \caption{Similar to Figure \ref{fig:AccretedMass_Zsun}, but for $Z=0.1\,Z_\odot$.}
     \label{fig:AccretedMass_0.1Zsun}
\end{figure*} 

Figure \ref{fig:AccretedMass_0.1Zsun} shows the allowed region for AICs of NSs at $Z=0.1\,Z_\odot$. Usually, lower metallicity results in weaker wind mass loss of He stars and final carbon-oxygen cores with higher mass, so that the initial masses of He stars as a final fate of NS would be decreased overall, which is clearly shown in Figure \ref{fig:NSMass}. He stars formed at lower-metallicty environments also tend to be more compact and, hence, be more difficult to undergo MT, compared with equal-mass He stars at higher metallicty. In comparison to that at solar metallicity, Figure \ref{fig:AccretedMass_0.1Zsun} reveals that the {{sub-solar metallicity}} AIC region for the same primary NS with a specific accretion rate moves toward a lower initial He star mass despite no significant change in the size of the region.

\begin{figure*}
     \centering
     \includegraphics[width=1.0\textwidth]{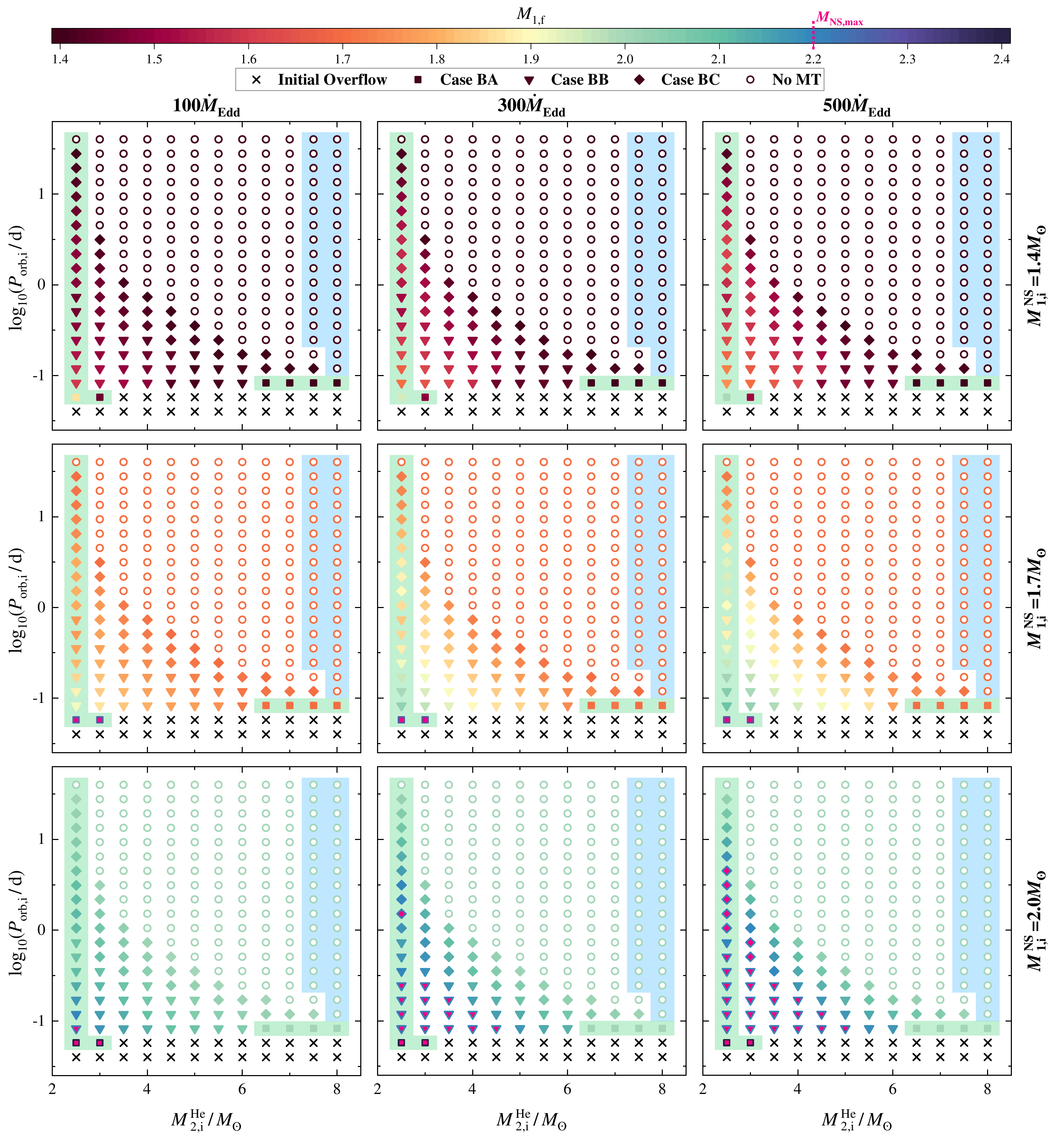}
     \caption{Similar to Figure \ref{fig:AccretedMass_Zsun}, but consider the accretion-induced magnetic field decay of NS.}
     \label{fig:AccretedMass_Zsun_B}
\end{figure*}

{{We now use an empirical formula from \cite{oslowski2011}, i.e., }}
\begin{equation}
\label{equ:BDecay}
    B_{\rm NS} = (B_{\rm NS,0} - B_{\rm NS,min})\exp(-{dM}/{\Delta M_{\rm d}})+B_{\rm NS,min},
\end{equation}
{{to consider possible accretion-induced magnetic field decay, where $B_{\rm NS,0}$ is the initial magnetic field which is listed in Section \ref{sec:properties} for different super-Eddington accretion rates, $B_{\rm NS,min} = 10^8\,{\rm G}$}} {{is the minimal magnetic field of an NS}}, $dM$ is the accreted mass, and $\Delta M_{\rm d}=0.05\,M_\odot$ given by \cite{oslowski2011}. {{By combining Equation (\ref{equ:SuperEddRate2}) and Equation (\ref{equ:BDecay}), we model the evolution of the NS magnetic field and accretion-induced with the accreted mass in \texttt{MESA}. By setting three different initial critical accretion rates including $100\dot{M}_{\rm Edd}$, $300\dot{M}_{\rm Edd}$, and $500\dot{M}_{\rm Edd}$, our simulated results at $Z=Z_\odot$ are displayed in Figure \ref{fig:AccretedMass_Zsun_B}.}} If we consider the possible accretion-induced magnetic field decay, Figure \ref{fig:AccretedMass_Zsun_B} demonstrates that NSs are unlikely to accrete enough materials to become mgBHs if the initial mass of primary NS is $M_{\rm 1,i}^{\rm NS}\lesssim1.7\,M_\odot$ since the magnetic field strength and accretion rate decay rapidly with the increase of the NS mass. AICs can only be achievable when the primaries of binary NS--He star systems are high-mass NSs with $M_{\rm 1,i}^{\rm NS}\gtrsim2.0\,M_\odot$ and the accretion rates are $\gtrsim300\,\dot{M}_{\rm Edd}$. However, the AIC regions still reduce significantly relative to those without consideration of accretion-induced magnetic field decay. As shown in Figure \ref{fig:AccretedMass_Zsun_B}, these AICs occur in very close-orbit binary systems with low-mass He stars which mainly undergo Case BB MT. Thus, if the magnetic field of NS could decay due to accretion, AICs could be hardly to happen via super-Eddington stable MT. 

\begin{figure*}
     \centering
     \includegraphics[width=0.667\textwidth]{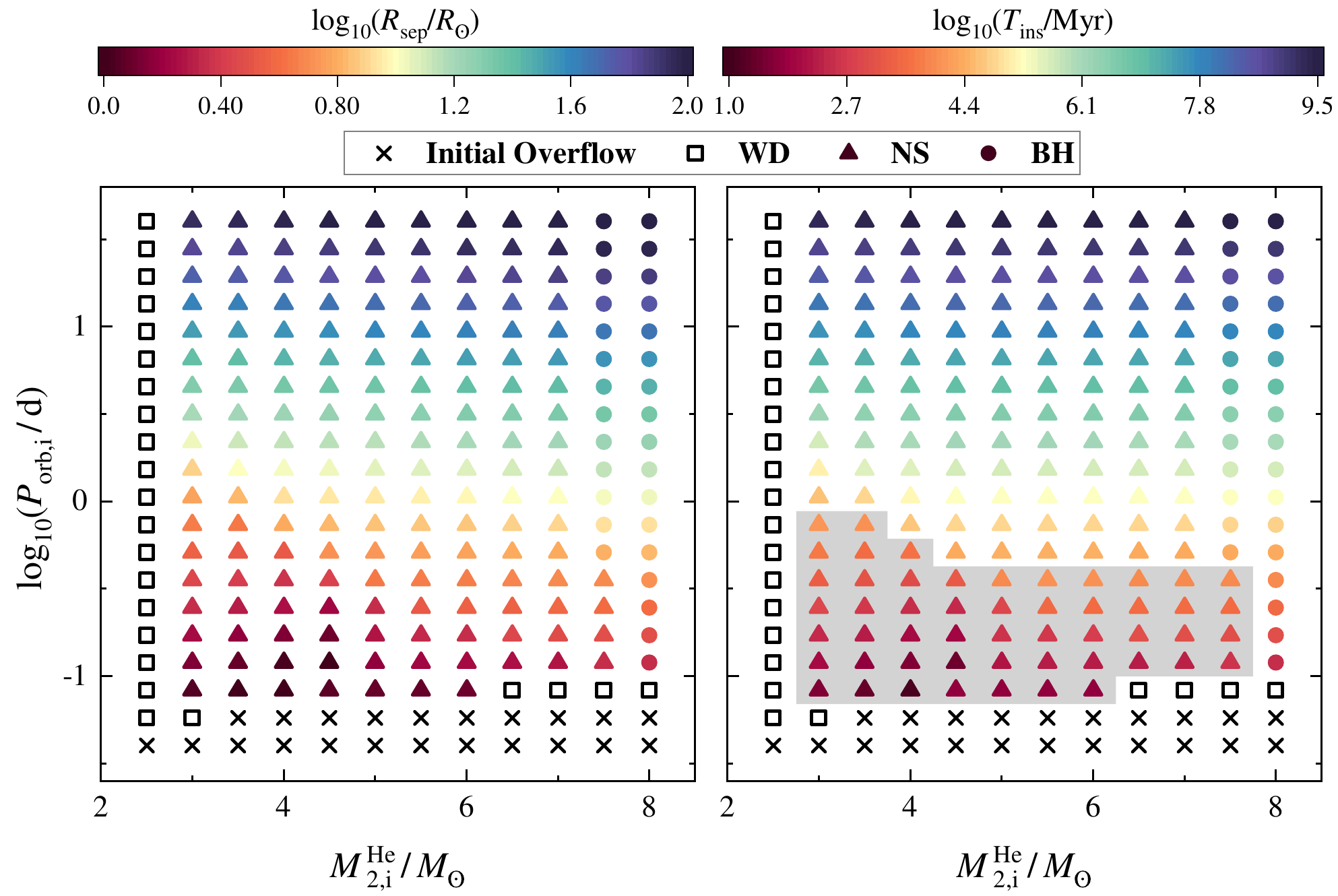}
     \caption{{{The left panel corresponds to the post-MT separation at the pre-SN stage and the right one refers to the insprial time of the massive BNS and mgBH--NS systems, respectively. The light gray region marks the parameter space in which the compact object binaries can merge within the Hubble time. Since different initial primary NS masses and Eddington accretion limits have little impact on post-ME separations and insprial times, $M^{\rm NS}_{1,i}=1.4\,M_\odot$ and an accretion limit of $100\dot{M}_{\rm Edd}$ are set here, while results by considering other different parameters limits are not shown. }}}
     \label{fig:PostMerger}
\end{figure*}

Accreting NSs during the Case BB/BC MT with super-Eddington accretion could form a population of mgBH--NS binaries. Because BHs formed after AICs of NSs could not continue to maintain super-Eddington accretions, the final masses of these mgBHs would be approach to or be a little higher than the Tolman-Oppenheimer-Vollkoff (TOV) TOV mass of NS. Furthermore, this formation channel could generate amount of massive BNS systems, in which the primary NSs could have a mass that is close to the TOV mass of NS, if AICs do not happen during Case BB/BC MT. {{In Figure \ref{fig:PostMerger}, we show the post-MT separations of the evolved systems and then further estimate the insprial time of the massive BNS and mgBH--NS systems. The SN kicks of our simulations are ignored, since}} most of second SNe from the systems that undergo Case BB/BC MT would be ultra-stripped SNe \citep{Tauris2015}. {{Our simplified simulations suggest that massive BNS and mgBH--NS systems which can merge within the Hubble time require an initial orbital period of $P_{\rm orb,i}\lesssim0.4-1\,{\rm d}$.}} \cite{Tauris2017} and \cite{Hu2023} present that most of BNS and NSBH binaries would be survived after the second ultra-stripped SNe due to weak kicks \cite[see discussions in][]{Tauris2015} and can merge within the Hubble time if $P_{\rm orb,i}\lesssim0.4\,{\rm d}$ {{which is consistent with our simplified simulations in Figure \ref{fig:PostMerger}}}. Therefore, it is expected that these high-mass BNS and mgBH--NS binaries formed through super-Eddington stable MT are ideal GW sources in LIGO and LISA bands.

\section{Discussion}

\subsection{EM Signatures}

It has long been proposed that disrupted BH--NS mergers can power gamma-ray bursts \citep[GRB; e.g.,][]{Paczynski1986,Paczynski1991,Eichler1989,Narayan1992A,Zhang2018,Gottlieb2023} and kilonova emissions \citep[e.g.,][]{li1998,Metzger2010,Kyutoku2015,Kawaguchi2016,Kasen2017,Barbieri2019,Zhu2020,Zhu2022Long,Darbha2021,Gompertz2023}. {{Whether an NS can be tidally disrupted by the primary BH and eject a certain amount of material to generate EM counterparts could be described by the total amount of remnant mass $M_{\rm total,fit}$ outside the BH horizon}} as a nonlinear function of $M_{\rm BH}$, the dimensionless projected aligned spin $\chi_{{\rm BH},z}$, $M_{\rm NS}$, and $R_{\rm NS}$\footnote{{{NS spin can affect tidal disruption probability of BH--NS mergers. He stars can be efficiently spun up via tidal effects during the MT stage and finally remain fast-spinning NSs after SN explosions \citep{Hu2023}. However, these NSs can possess high magnetic field strengths, causing them to rapidly lose rotational energy through spin-down processes. Therefore, the spins of NSs when BH-NS mergers may always have little effect on tidal disruption.}}}, which has been given by an empirical model from \cite{Foucart2018}, i.e.,
\begin{equation}
\label{equ:TotalEjectaMassFunction}
    \frac{M_{\rm total,fit}}{M^{\rm b}_{\rm NS}} = \left[\max\left(\alpha\frac{1 - 2C_{\rm NS}}{\eta^{1 / 3}} - \beta \widetilde{R}_{\rm {ISCO}}\frac{C_{\rm NS}}{\eta} + \gamma , 0 \right)\right]^ {\delta},
\end{equation}
where $M_{\rm NS}^{\rm b}$ is the NS baryonic mass, $C_{\rm NS} = GM_{\rm NS}/c^2R_{\rm NS}$ is the compactness, $\widetilde{R}_{\rm ISCO}= 3 + Z_2 - {\rm {sign}}(\chi_{{\rm BH},z})\sqrt{(3 - Z_1)(3 + Z_1 + 2Z_2)}$ is the normalized radius of the BH innermost stable circular orbit with $Z_1 = 1 + (1 - \chi_{{\rm BH},z}^2) ^ {1 / 3} [(1 + \chi_{{\rm BH},z})^{1 / 3} + (1 - \chi_{{\rm BH},z})^{1 / 3}]$ and $Z_2 = \sqrt{3 \chi_{{\rm BH},z}^2 + Z_1^2}$, $\eta = Q / (1 + Q) ^ 2$ in Equation (\ref{equ:TotalEjectaMassFunction}), and $Q = M_{\rm BH} / M_{\rm NS}$ is the mass ratio. The fitting parameters include $\alpha = 0.406$, $\beta = 0.139$, $\gamma = 0.255$, and $\delta = 1.761$. BH--NS mergers with system parameters falling into the parameter space where $M_{\rm total,fit} = 0$ correspond to plunging events, {{while those with $M_{\rm total,fit} > 0$ can allow tidal disruption to generate EM signals}}.

\begin{figure}
    \centering
    \includegraphics[width = 1\linewidth , trim = 50 35 105 30, clip]{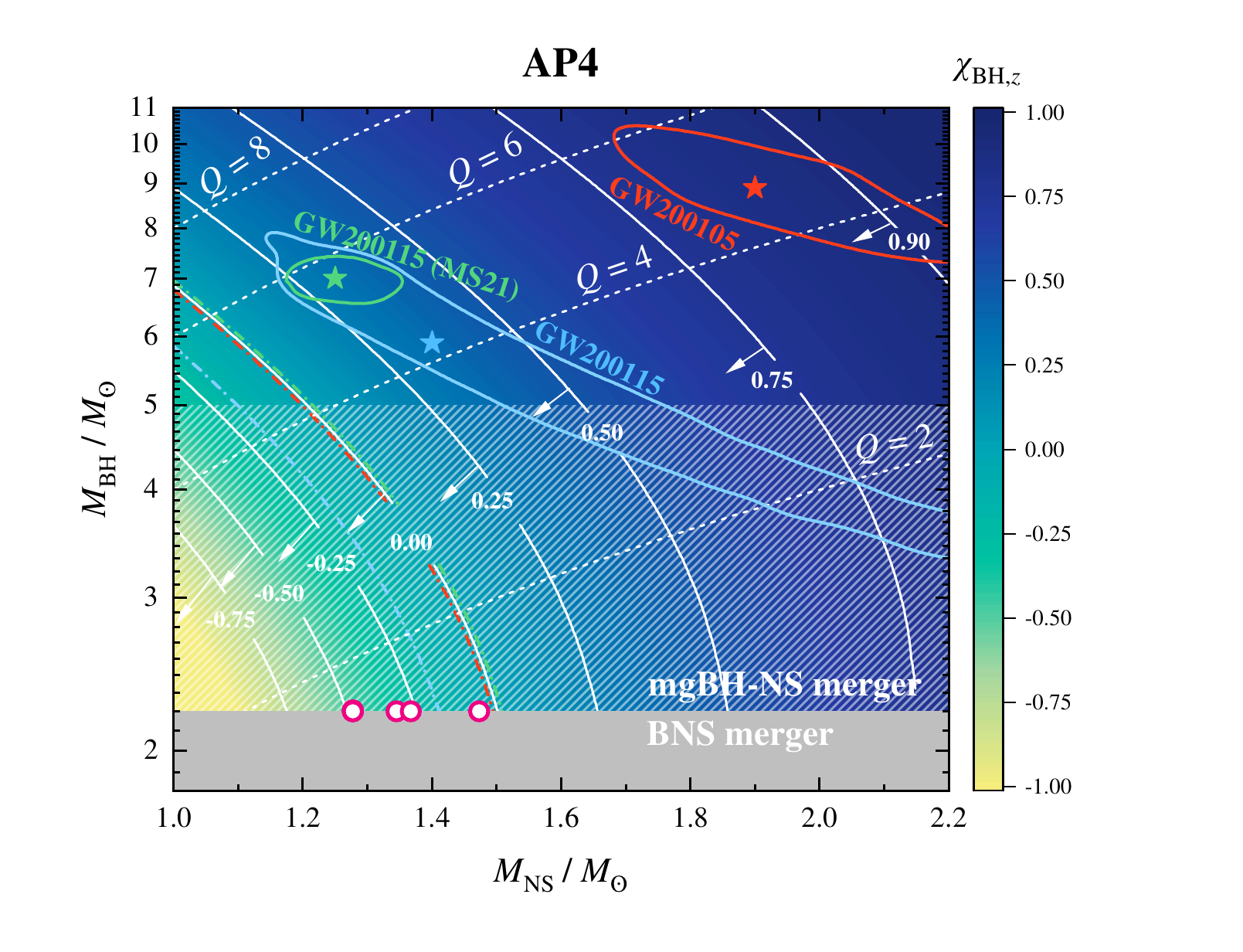}
    \caption{The source-frame mass parameter space for BH--NS merger systems to allow tidal disruption of the NS by the BH with consideration of the AP4 EoS model. The dashed lines represent mass ratio from $Q = 2$ to 8. We mark several values of the primary BH spin along the orbital angular momentum from $\chi_{{\rm BH},z}=-0.75$ to 0.90 as solid lines. For a specific $\chi_{{\rm BH},z}$, the BH--NS mergers with component masses located at the bottom left parameter space (denoted by the direction of the arrows) can {{have $M_{\rm tot,fit}>0$, indicating that these mergers can}} allow tidal disruption. {{The gray and white shadow regions represent that the mass of the primary compact objects falls below the NS maximum mass and into the mass gap, respectively. The binary systems that can form mgBH--NS mergers, as shown in Figure \ref{fig:AccretedMass_Zsun}, are marked as pink empty points.}} The $90\%$ credible posterior distributions (colored solid lines) and the medians (colored stars) of GW200105 (orange), GW200115 \citep[blue;][]{Abbott2021BHNS}, and GW200115 (green) obtained by applying an astrophysically motivated priors \citep{Mandel2021} are displayed, while corresponding median values of $\chi_{{\rm BH},z}$ for these two sources are marked as dashed-dotted lines. }
    \label{fig:TidalDisruptionRegion}
\end{figure}

We adopt the AP4 model \citep{Akmal1997} as the NS EoS, whose TOV mass is $M_{\rm TOV}=2.22\,M_\odot$ nearly consistent with $M_{\rm NS,max}$ we used in Section \ref{sec:model}. AP4 is one of the most likely EoSs allowed by the constraint on the observations of GW170817 \citep[e.g.,][]{Abbott2018GW170817}. We calculate $M_{\rm NS}^{\rm b}$ as an empirical function of $M_{\rm NS}$ given by \cite{Gao2020}, i.e., $M_{\rm NS}^{\rm b}=M_{\rm NS} + A_1\times M_{\rm NS}^2+A_2\times M_{\rm NS}^3,$ where $A_1=0.045$ and $A_2=0.023$, and $M_{\rm NS}^{\rm b}$ and $M_{\rm NS}$ in this function are in units of $M_\odot$. $C_{\rm NS}\approx 1.1056\times(M_{\rm NS}^{\rm b}/M_{\rm NS} - 1)^{0.8277}$ is estimated following the empirical formula constructed by \cite{Coughlin2017}. In Figure \ref{fig:TidalDisruptionRegion}, we display the tidal disruption region for BH--NS mergers dependent on $M_{\rm BH}$, $\chi_{{\rm BH},z}$, and $M_{\rm NS}$ {{by using Equation (\ref{equ:TotalEjectaMassFunction})}}. 

For the standard isolated binary formation channel of BH--NS mergers, {{BH component unusually forms firstly in a wide orbit and}}, hence, could have a negligible spin since the tides are too weak to spin its progenitor up \cite[see Figure 1 in][]{Qin2018}\footnote{{{Under extreme conditions, the NS in a BH--NS binary could be born first due to a reversal in the mass ratio of progenitor stars during the Case A MT stage \citep[][]{Pols1994,Jaime2021,Broekgaarden2021BHNS,Mould2022,Adamcewicz2023}. The BH progenitors can thus be tidally spun up after the common-envelope stage, finally remaining fast-spinning BHs \citep{Hu2022BHNS}. }}}. As shown in Figure \ref{fig:TidalDisruptionRegion}, if the mass gap is intrinsically existent, disrupted events could only contribute a limited fraction of BH--NS population in the universe, which would require a mass space of $M_{\rm BH}\lesssim6\,M_\odot$ and $M_{\rm NS}\lesssim1.2\,M_\odot$. For examples, GW200105 (GW200115) reported by the LVK Collaboration in O3 \citep{Abbott2021BHNS} is formed through a merger between a $8.9^{+1.1}_{-1.3}\,M_\odot$ ($5.9^{+1.4}_{-2.1}\,M_\odot$) BH with a dimessionless projected aligned spin of $-0.01^{+0.10}_{-0.16}$ ($-0.18^{+0.21}_{-0.57}$) and a $1.9^{+0.2}_{-0.2}\,M_\odot$ ($1.4^{+0.6}_{-0.2}\,M_\odot$) NS, inferred from low-spin priors of the secondary. \cite{Mandel2021} found that GW200115 could be more constrained, with $M_{\rm BH} = 7.0^{+0.4}_{-0.4}\,M_\odot$ and $M_{\rm NS} = 1.25^{+0.09}_{-0.09}\,M_\odot$, by adopting a zero-spin prior preference. Thus, these two BH--NS mergers could possess BH components with mass above the mass gap, and their formation channels could align with the standard picture \citep[e.g,][]{Broekgaarden2021,Chattopadhyay2022,Zhu2022Population,Jiang2023}. Apparently, Figure \ref{fig:TidalDisruptionRegion} suggested that these mergers could hardly make tidal disruption to generate EM counterparts, consistent with the conclusion from \cite{Zhu2021No,Fragione2021,Gompertz2022,DOrazio2022,Biscoveanu2023}. {{In contrast, mgBHs could be more likely to tidally disrupt more massive NSs. Figure \ref{fig:TidalDisruptionRegion} shows that mergers between zero-spin mgBHs and NSs can easily undergo tidal disruption as long as $M_{\rm NS}\lesssim1.2-1.5\,M_\odot$.}}

{{Our detailed binary simulations shown in Section \ref{sec:properties} suggest that mgBHs formed via super-Eddington accretion during the Case BB/BC MT could have a mass nearly equal to or slightly higher than $M_{\rm NS,max}$. Unless the mass of secondaries is $\gtrsim1.5\,M_\odot$, tidal disruption can always happen.}} {{As an example in Figure \ref{fig:TidalDisruptionRegion}, we show all our \texttt{MESA} grid data capable of forming mgBH--NS mergers at a metallicity of $Z_\odot$ in Figure \ref{fig:AccretedMass_Zsun}. Although only four data points of mgBH--NS mergers are visible in Figure \ref{fig:TidalDisruptionRegion}, we note that actually a total of 82 data points are included, as they are overlapping. One can find that}} {{ the secondaries for those binary systems that experience Case BB/BC could mostly have a mass of $\lesssim1.5\,M_\odot$}}{{, which are located in the parameter region that allows tidal disruption}}.  Thus, it is expected that a large proportion of mgBH--NS mergers formed through super-Eddington stable MT would occur tidal disruption to generate EM signals.

\subsection{Constraints on the NS TOV Mass and EoS} 

In Section \ref{sec:properties}, we suggest that binary systems undergoing super-Eddington stable MT can finally generate a population of GW NS binaries, comprising either an NS with a mass close to $M_{\rm TOV}$ or a mgBH with a mass slightly higher than $M_{\rm TOV}$. Distinguishing between these high-mass BNS and mgBH-NS mergers using GW and EM observations can help constrain the NS TOV mass and EoS.

In the case of a mgBH--NS mergers, GW emissions would abruptly terminate upon the tidal disruption of the NS, without exciting quasi-normal modes \citep{Shibata2009}. However, for a massive BNS involving a $M_{\rm TOV}$ NS, the outcome of the merger is a prompt collapse accompanied with ringdown emission from the remnant BH. The difference of GW merger phase in the kilohertz range between these scenarios could be discerned with the advent of third-generation detectors \citep[e.g.,][]{Shibata2009,Kyutoku2020}.

Despite prompt collapse after merging, a typical-mass NS in an asymmetric BNS system with a companion of a $M_{\rm TOV}$ NS can be tidally disrupted to power bright kilonova signal. These asymmetric BNS mergers can generate disk outflow and massive lanthanide-rich dynamical ejecta, whose kilonovae are similar to those from mergers between $\gtrsim5\,M_\odot$ BH and NS \citep[e.g.,][]{Sekiguchi2016,Bernuzzi2020}. However, numerical relativity simulations revealed that the mass of dynamical ejecta from mgBH--NS mergers could be lower by an order of magnitude compared to that from corresponding BNS binaries, or even negligible \citep{Foucart2019,Kyutoku2020,Hayashi2021}. Thus, the kilonova emissions from low-mass mgBH--NS mergers could lack the contribution from lanthanide-rich dynamical ejecta, which can be used to distinguish mgBH--NS mergers from high-mass BNS mergers through follow-up observations.

Thus, we expect that NS-He star binaries with super-Eddington stable MT could serve as a plausible formation channel to form a population of high-mass BNS and low-mass mgBH--NS GWs. Future GW and EM observations on a large number of these NS binaries could provide a precise mass boundary between NSs and BHs.

\section{Conclusions}

In this paper, we propose a new scenario for the formation of mgBH--NS binaries. {{A fraction of}} NSs with companions of He stars could have super-Eddington stable MT, as suggested by both observations and theory. We evolve NS--He star binaries by our detailed binary simulations to explore the parameter space that allows for the AICs of NSs. Since super-Eddington accretion can be maintained during the Case BB/BC MT stage, our simulated results reveal that AIC events tend to happen when the primaries NS have an initial mass $\gtrsim1.7\,M_\odot$ with an accretion rate of $\gtrsim300\,\dot{M}_{\rm Edd}$. This formation scenario can thus generate a population of NS binaries consisting of a mgBH with a mass slightly higher than $M_{\rm TOV}$ for AIC events or an NS with a mass close to $M_{\rm TOV}$ if AICs do not happen, which can eventually merge within the Hubble time to become ideal GW sources. {{However, it is difficult to evaluate how common such a scenario is for forming mgBH--NS mergers based on our present knowledge. On the one hand, low-mass NSs are more common than high-mass NSs inferred by the Galactic pulsar observations \citep{Lattimer2012,Antoniadis2016,Alsing2018,Farr2020} and population synthesis simulations \citep[e.g.,][]{Giacobbo2018,VignaGomez2018,Broekgaarden2022}. The lack of high-mass primary NSs in BNS systems makes it difficult for AICs to happen. On the other hand, GW190425 was measured to have a heavy primary NS with a mass of $\sim1.61-2.52\,M_\odot$ \citep{gw190425}, accompanied by an unexpectedly high event rate density, if it had a BNS origin. Furthermore, the NSs observed in BNS and BH--NS mergers detected via GWs exhibit a uniform distribution in mass \citep{Landry2021,Zhu2022Population,Abbott2023Population}, contrary to the Galactic pulsar observations and population synthesis simulations. Systems between high-mass NSs and He stars may be widespread in the universe, suggesting that AICs via super-Eddington stable MT are not necessarily impossible.}}

{{These mgBH--NS mergers can easily lead to tidal disruption, generating bright EM signals.}} Future GW and EM observations on the population of these NS binaries formed via super-Eddington stable MT could help us constrain the TOV mass and EoS of NSs. 

Most recently, \cite{ligo2023} reported a GW compact binary merger candidate in the fourth observing run, S230529ay, which was observed solely by the LIGO Livingston Observatory. S230529ay could be a high-confidence GW candidate with a false alarm rate of $1$ per $160.44\,{\rm yr}$ estimated by the online analysis. The probabilities of classifying this GW signal are $62\%$ and $31\%$ for BH--NS and BNS merger origins, respectively. Under the assumption that the source had an astrophysical origin, the probability that the mass of the primary compact object lied in the mass range of the lower mass gap is $98.5\%$, while the probability that the secondary compact object was an NS is $>99\%$. Based on present limited information on this GW, we could suspect that the primary of this GW could not be much higher than the TOV mass. Thus, S230529ay could be a GW candidate of mgBH--NS binary merger, and could possibly originate from super-Eddington stable MT formation channel.

\section*{Acknowledgements}

JPZ thank Team COMPAS group for useful discussions. This work was supported by Anhui Provincial Natural Science Foundation (grant No. 2308085MA29) and the Natural Science Foundation of Universities in Anhui Province (grant No. KJ2021A0106).

\section*{Data Availability}

The data generated in this work will be shared upon reasonable request to the corresponding author.

\bibliographystyle{mnras}
\bibliography{ms} 




\bsp	
\label{lastpage}
\end{document}